\documentclass[aip,apl,amsfonts,amssymb,amsmath,groupedaddress,twocolumn]{revtex4}

\usepackage{graphicx}
\usepackage{amsmath}
\usepackage{amsfonts}
\usepackage{amssymb}
\usepackage{graphicx}
\usepackage{epstopdf}

\begin{document}
\title{A Simple Boltzmann Transport Equation for Ballistic to Diffusive Transient Heat Transport}

\author{Jesse Maassen}
\email{jmaassen@purdue.edu}
\author{Mark Lundstrom}
\affiliation{Network for Computational Nanotechnology, School of Electrical and Computer Engineering, Purdue University, West Lafayette, IN 47907, USA}

\begin{abstract}
Developing simplified, but accurate, theoretical approaches to treat heat transport on {\it all} length and time scales is needed to further enable scientific insight and technology innovation. Using a simplified form of the Boltzmann transport equation (BTE), originally developed for electron transport, we demonstrate how ballistic phonon effects and finite-velocity propagation are easily and naturally captured. We show how this approach compares well to the phonon BTE, and readily handles a full phonon dispersion and energy-dependent mean-free-path. This study of transient heat transport shows {\it i)} how fundamental temperature jumps at the contacts depend simply on the ballistic thermal resistance, {\it ii)} that phonon transport at early times approach the ballistic limit in samples of any length, and {\it iii)} perceived reductions in heat conduction, when ballistic effects are present, originate from reductions in temperature gradient. Importantly, this framework can be recast {\it exactly} as the Cattaneo and hyperbolic heat equations, and we discuss how the key to capturing ballistic heat effects is to use the correct physical boundary conditions.   
\end{abstract}


\maketitle

\section{Introduction}
Transient thermal transport is a problem of great interest from both fundamental and applied perspectives.  For example, it is an important factor in self-heating in small electronic devices, in phase-change memory and in heat-assisted magnetic recording \cite{Cahill2014}.  Time/frequency-domain thermal reflectance (TDTR/FDTR), make use of rapid thermal transients to measure the thermal properties of materials \cite{Koh2007,Regner2013}. It is now well known that ballistic phonon transport becomes important in structures with small feature sizes as well as in large structures under rapid transient conditions \cite{Minnich2011,Wilson2014}. While first principles simulations \cite{Sellan2010c,Esfarjani2011,Yang2013,Bae2013}, and other physically detailed techniques, have contributed to our understanding of the basic science, there remains a need for simplified thermal transport approaches that capture the essential physics and that are computationally tractable. A technique that can be derived from the phonon Boltzmann transport equation (BTE) with clearly identified simplifications and that provides reasonable accuracy and excellent computational efficiency all the way from the ballistic to diffusive limits is described in this paper.

Traditionally, heat transport has been described by Fourier's law with the heat equation (HE), but an unphysical implication of this approach is that phonons can travel at infinite speed \cite{Joshi1993}. The hyperbolic heat equation (HHE) resolves this issue by adding a term to the heat equation that ensures a finite propagation velocity \cite{Joshi1993,Chen2002}. It is generally understood that these approaches are valid only when heat transport is diffusive and the characteristic length scales are much larger than the phonon MFP \cite{ChenBook}. Sub-continuum approaches such as the BTE \cite{Mahan1988,Claro1989,Majumdar1993,Joshi1993,Chen2002,Narumanchi2006,Escobar2008,Sellan2010a,Hua2014,Regner2014,Vermeersch2014a,Vermeersch2014b}, molecular-dynamics \cite{Gomes2006,Hu2009,Sellan2010b,Sullivan2013} and Monte-Carlo simulations \cite{Mazumder2001,Lacroix2005,Mittal2010}, capture quasi-ballistic phonon transport as well as retardation effects due to finite-velocity propagation. It is generally thought that the treatment of ballistic phonons requires going beyond the HE or HHE \cite{Joshi1993}.

Sophisticated detailed modeling has added much to the fundamental understanding of ballistic effects in phonon transport, but there is also a need for simplified, computationally efficient, but accurate analysis techniques.  Several such approaches have been proposed \cite{Mahan1988,Claro1989,Majumdar1993,Joshi1993,Chen2002,Alvarez2007,Maznev2011,Wilson2013,Hua2014,Regner2014,Vermeersch2014a,Vermeersch2014b}. These often follow one (or both) of the following strategies: i) use the phonon BTE simplified to obtain solutions for one specific problem, ii) use a two-channel model to treat ballistic and diffusive phonons separately. Our goal in this paper is to contribute to this work by introducing a technique that combines the physics of the phonon Boltzmann equation with the computational efficiency of diffusive equations.

In a recent paper, we showed that the McKelvey-Shockley flux method, a simple form of the BTE, provides highly accurate solutions for steady-state thermal transport from the ballistic to diffusive limits \cite{Maassen2014}.  In this paper, we extend the work in \cite{Maassen2014} to the transient case and demonstrate that it naturally includes ballistic and diffusive phonon transport and finite-velocity heat propagation, and thus is applicable on all length and time scales. Good agreement with the full phonon BTE is obtained, while requiring substantially less computational effort. Interestingly, we prove that our simple BTE can be rewritten exactly as the Cattaneo and hyperbolic heat equations.  Both ballistic effects and the finite propagation time are treated when the correct boundary conditions are used.

The paper is organized as follows: Section \ref{sec:theory} describes the McKelvey-Shockley flux method and extends the work in \cite{Maassen2014} to the transient case.  In Section \ref{sec:example}, we demonstrate the technique by treating transient heat conduction through a dielectric film. Within this section, extensions to realistic phonon dispersions and energy-dependent mean-free-paths are discussed and demonstrated, and the technique is benchmarked against numerical solutions to the full phonon BTE.  Section \ref{sec:hhe} describes the connection between the McKelvey-Shockley flux method and well-known traditional heat transport equations, and lastly, in Section \ref{sec:summary}, we summarize our findings.

\section{Theoretical Approach} 
\label{sec:theory}
\subsection{Phonon flux and density}
\label{sec:theory_flux}
Our starting point is the McKelvey-Shockley flux method \cite{Mckelvey1961,Shockley1962}, an approach developed to treat electron transport and recently reformulated to handle phonon/heat transport \cite{Maassen2014}. In this work, we consider 1D transport along $x$, with $y$ and $z$ directions extending to infinity. This technique categorizes phonons into two components, those that are forward moving ($v_x>0$) and backward moving ($v_x<0$). The governing equations describe how the phonon flux, the product of phonon density and phonon velocity, vary in space and time. The McKelvey-Shockley flux equations are \cite{Alam1995}:
\begin{align}
\frac{1}{v_x^+}\frac{{\rm d} F^+}{{\rm d} t} + \frac{{\rm d} F^+}{{\rm d} x} &= -\frac{F^+}{\lambda}+\frac{F^-}{\lambda}, \label{mk_flux1} \\
-\frac{1}{v_x^+}\frac{{\rm d} F^-}{{\rm d} t} + \frac{{\rm d} F^-}{{\rm d} x} &= -\frac{F^+}{\lambda}+\frac{F^-}{\lambda}, \label{mk_flux2}
\end{align}
where $F^{+/-}(x,t,\epsilon)$ are the forward/backward phonon fluxes [units: \#phonons  m$^{-2}$s$^{-1}$eV$^{-1}$], $v_x^+(\epsilon)$ is the average $x$-projected phonon velocity, $\lambda(\epsilon)$ is the mean-free-path for backscattering, and $\epsilon$ is the phonon energy. Eqns. (\ref{mk_flux1})-(\ref{mk_flux2}) are coupled through their scattering terms (right-hand terms), which describe how phonons scatter in/out of each flux type. Scattering is described by $\lambda$, which is the average distance traveled along $x$ before a forward-moving (`$+$') or backward-moving (`$-$') phonon is backscattered \cite{Jeong2010}. $v_x^+$ is responsible for finite-velocity and retardation effects. We refer readers to \cite{footnote_velo_mfp} for the definitions of $\lambda$ and $v_x^+$. The McKelvey-Shockley flux method can be derived from the BTE; the case of steady-state is treated in \cite{Alam1993,Rhew2002}, and a more general derivation will be presented in future work.

\begin{figure}	
\includegraphics[width=8.5cm]{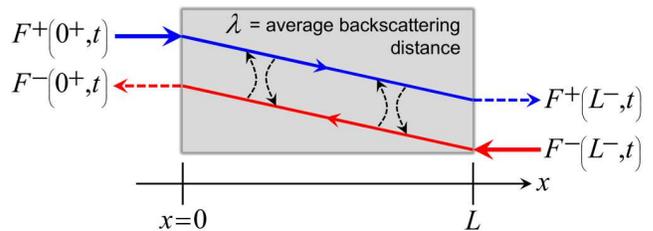}
\caption{Thermal conductor of length $L$. The mean-free-path for backscattering $\lambda$ controls the scattering between the forward/backward fluxes. By specifying the injected phonon fluxes (solid arrows), the McKelvey-Shockley flux equations describe the evolution of the fluxes inside the material.} \label{fig1}
\end{figure}

Since Eqns. (\ref{mk_flux1})-(\ref{mk_flux2}) are first order in $x$ and $t$, we must specify one spatial and temporal boundary condition for each flux component. The spatial boundary conditions correspond to specifying the injected phonon fluxes at the boundaries: $F^+(0^+,t,\epsilon)$ at the left end, and $F^-(L^-,t,\epsilon)$ at the right end, where $L$ is the length of the thermal conductor (depicted in Fig. \ref{fig1}). The temporal boundary conditions, depending on the problem at hand, will generally correspond to specifying the forward/backward fluxes at some given time $t'$: $F^+(x,t',\epsilon)$ and $F^-(x,t',\epsilon)$.

The directed fluxes $F^{\pm}$ are related to the net phonon flux $F$ and the phonon density $n$ through the relations:
\begin{align}
F(x,t,\epsilon) &= F^+(x,t,\epsilon)-F^-(x,t,\epsilon) \label{net_flux}, \\
n(x,t,\epsilon) &= \frac{F^+(x,t,\epsilon)+F^-(x,t,\epsilon)}{v_x^+(\epsilon)}. \label{ph_dens}
\end{align}
The total net phonon flux $F^{\rm tot}$ and total phonon density $n^{\rm tot}$ are obtained by integrating over energy:
\begin{align}
F^{\rm tot}(x,t) &= \int_0^{\infty}\!\!F(x,t,\epsilon)\, {\rm d}\epsilon \label{tot_net_flux}, \\
n^{\rm tot}(x,t) &= \int_0^{\infty}\!\!n(x,t,\epsilon)\, {\rm d}\epsilon \label{tot_ph_dens}.
\end{align}

Eqns. (\ref{mk_flux1})-(\ref{mk_flux2}), with the appropriate boundary conditions, describe the spatial and temporal evolution of phonons in a material. Next, we show how to relate phonon fluxes to quantities relevant for heat transport.

\subsection{Heat current and density}
\label{sec:theory_heat_flux}
The heat current is obtained by simply multiplying the phonon flux times the phonon energy:
\begin{align}
I_Q^{\pm}(x,t,\epsilon) &= \epsilon \, F^{\pm}(x,t,\epsilon). \label{def_heat_curr}
\end{align} 
We can then define the net heat current ($I_Q$) and heat density ($Q$) as:
\begin{align}
I_Q(x,t,\epsilon) &= I_Q^+(x,t,\epsilon) - I_Q^-(x,t,\epsilon), \label{def_net_heat_curr} \\
Q(x,t,\epsilon) &= \frac{I_Q^+(x,t,\epsilon) + I_Q^-(x,t,\epsilon)}{v_x^+(\epsilon)}. \label{def_heat_den}
\end{align} 
 
By integrating over energy, and hence all phonons, we obtain the total heat current ($I_Q^{\rm tot}$) and heat density ($Q^{\rm tot}$):
\begin{align}
I_Q^{\rm tot}(x,t) &= \int_0^{\infty} I_Q(x,t,\epsilon)\,{\rm d}\epsilon,\;\;\;\text{[W\,m$^{-2}$]} \label{def_heat_curr_tot} \\
Q^{\rm tot}(x,t) &= \int_0^{\infty}Q(x,t,\epsilon)\,{\rm d}\epsilon,\;\;\;\text{[J\,m$^{-3}$]} \label{def_heat_den_tot}
\end{align}

By replacing phonon flux $F^{\pm}$ by heat current $I_Q^{\pm}$, all the equations presented in Section \ref{sec:theory_flux} can be used to treat heat transport. From this point on, unless deemed important, the explicit dependence of quantities on energy $\epsilon$ will be dropped to simplify the presentation. Keep in mind that a final integration over energy is required to compute the total phonon flux/density and total heat current/density.

\subsection{Temperature} 
\label{sec:theory_temp}
In addition to heat current and heat density, it is common to calculate temperature profiles. We note, however, that temperature is an equilibrium quantity, and in cases where the phonon population is far from equilibrium its definition may become ambigious. In this work, we will consider small applied temperature differences at the ends of the thermal conductor ($\Delta T = T_L - T_R$, where $T_{L,R}$ are the temperatures of the left/right contacts), which ensures that the system is near equilibrium and temperature is well defined. Note, however, that the McKelvey-Shockley flux approach itself is not strictly limited to small applied temperature differences. 

The forward/backward heat currents can be expanded as:
\begin{align}
I_Q^{\pm}(x,t) &= I_{Q,{\rm eq}}^{\pm} + \delta I_Q^{\pm}(x,t), \label{heat_curr_expand}
\end{align}
where $I_{Q,{\rm eq}}^+=I_{Q,{\rm eq}}^-$ is the equilibrium heat current resulting from a constant background reference temperature $T_{\rm ref}$, and $\delta I_Q^{\pm}$ is a small correction resulting from a small applied $\Delta T$. 

Temperature is related to heat density by $\delta Q = C_V \,\delta T$, where $C_V$ is the volumetric heat capacity. Using this with Eq. (\ref{def_heat_den}) and Eq. (\ref{heat_curr_expand}), we find that temperature can be expressed as:
\begin{align}
T(x,t) &= \left[ \frac{\delta T^+(x,t)+\delta T^-(x,t)}{2} \right] + T_{\rm ref}, \label{def_temp} \\ 
\delta T^{\pm}(x,t) &= \frac{2\,\delta I_Q^{\pm}(x,t)}{C_V \, v_x^+}, \label{def_temp_pm}
\end{align}
where $\delta T^{\pm}$ is the correction in temperature relative to $T_{\rm ref}$ for each phonon component (forward and backward) and $C_V$ is the heat capacity at $T_{\rm ref}$. 

The calculational procedure is as follows. Given some contact temperatures $T_L$/$T_R$ (which can be time-dependent), the injected heat fluxes from each contact are computed (the relation between $T_L$/$T_R$ and $I_Q^+(0^+)$/$I_Q^-(L^-)$ is shown later). Using the injected heat fluxes as boundary conditions, we calculate the directed heat fluxes for all $x$ and $t$ using Eqns. (\ref{mk_flux1})-(\ref{mk_flux2}), replacing $F^{\pm}$ with $I_Q^{\pm}$ (it is convenient to solve directly for $\delta I_Q^{\pm}$ since $I^{\pm}_{Q,{\rm eq}}$ only adds a constant). If a temperature profile is desired, $T(x,t)$ can be determined from Eqns. (\ref{def_temp})-(\ref{def_temp_pm}).

As we will illustrate in the next section, the McKelvey-Shockley flux method captures ballistic and finite-velocity propagation effects, and is thus applicable on all length and time scales. Later we will find that the McKelvey-Shockley equations are exactly equivalent to well-known diffusive equations.

\begin{figure}	
\includegraphics[width=6.5cm]{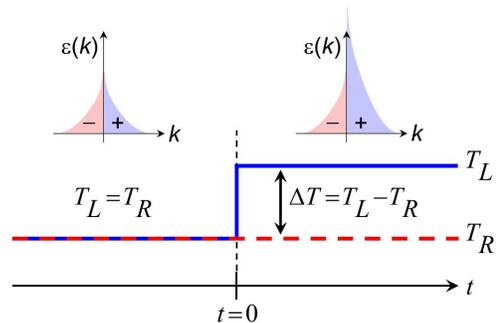}
\caption{(Color online) Applied temperature at the left and right contacts, $T_L$ and $T_R$, versus time. Top insets: area plots showing occupation of the forward-moving (blue) and backward-moving (red) phonon states injected at the left and right ends, respectively, of the thermal conductor. For $t<0$, $T_L=T_R$ leading to equal numbers of injected phonons. For $t>0$, $T_L = T_R + \Delta T$, which results in a larger number of forward states injected at the left side of the material. } \label{fig2}
\end{figure}

\section{Example: transient heat conduction across a dielectric film}
\label{sec:example}
Here we apply the McKelvey-Shockley flux method to an example of transient thermal conduction through a dielectric film. We assume the dielectric film of length $L$ (in which the electronic contribution to thermal transport can be neglected) is joined by two {\it ideal} thermalizing left and right contacts. As discussed in \cite{Maassen2014}, an {\it ideal} contact is in thermal equilibrium described by Bose-Einstein statistics and is perfectly absorbing/reflectionless, meaning phonons that reach the contact do not backscatter at the interface. Note that ideal contacts are assumed for this example, but this is not a limitation of the McKelvey-Shockley flux method; any specified injected heat currents at the ends of the thermal conductor can be used as the boundary conditions.

Before time $t=0^-$ the film is at equilibrium $T_{\rm ref}=300\,{\rm K}$, after $t=0^+$ the left contact temperature is raised by some small $\Delta T$ and the right contact temperature is kept fixed, as depicted in Fig. \ref{fig2}. In the case of equilibrium contacts, we can easily relate the contact temperature to the injected heat current:
\begin{align}
I_Q^+(0^+,t,\epsilon) &= \epsilon \, \frac{v_x^+(\epsilon)\,D(\epsilon)}{2}\,f_{\rm BE}(\epsilon,T_L(t)), \label{inj_IQ_1} \\
I_Q^-(L^-,t,\epsilon) &= \epsilon \, \frac{v_x^+(\epsilon)\,D(\epsilon)}{2}\,f_{\rm BE}(\epsilon,T_R(t)), \label{inj_IQ_2} 
\end{align} 
where $D(\epsilon)$ is the phonon density of states, $f_{\rm BE}$ is the Bose-Einstein distribution and the factor of two comes from half the states being forward and backward moving. Note that the product $v_x^+(\epsilon)\,D(\epsilon)/2 = M(\epsilon)/h$, where $M(\epsilon)$ is the phonon distribution of modes and $h$ is Planck's constant, which can be efficiently calculated from the phonon dispersion \cite{Jeong2011,Maassen2013a,Maassen2013b}. 

When $T_L=T_R$, the injected fluxes are equal and no net heat current flows. However, if $T_L >T_R$, the left contact will inject a larger heat current than the right contact which will drive a net heat current, as illustrated in Fig. \ref{fig2}. Using Eqns. (\ref{inj_IQ_1})-(\ref{inj_IQ_2}) as boundary conditions, we can solve the McKelvey-Shockley flux equations by replacing $F^{\pm}$ with $I_Q^{\pm}$ in Eqns. (\ref{mk_flux1})-(\ref{mk_flux2}). For a small $\Delta T=T_L-T_R$, we can expand Eq. (\ref{inj_IQ_1}) to first order around $T_R$, and solve the McKelvey-Shockley flux equations directly for $\delta I_Q^{\pm}$, as defined in Eq. (\ref{heat_curr_expand}).

\subsection{Diamond film: constant velocity and mean-free-path}
\label{sec:diamond}
In this example, we consider diamond as our dielectric material, so we may compare to the results of the full phonon BTE for the same problem \cite{Joshi1993}. Here, we use $\lambda= 60\,{\rm nm}$ \cite{footnote_fit_mfp}, which provides a good fit to the steady-state and transient results in Ref. \cite{Joshi1993}. Typically $\lambda(\epsilon)$ in a bulk material can vary by orders of magnitude, however in Ref. \cite{Joshi1993}, and thus in this work, an effective single energy-independent $\lambda$ is used.  

Given diamond's large Debye temperature ($\sim 1860 \,{\rm K}$) \cite{Joshi1993}, this material can be modeled at $300\,{\rm K}$ by assuming a linear phonon dispersion with group velocity $v_g = 12\,288 \,{\rm m/s}$. The distribution of modes in this case can be written as $M(\epsilon) = 3\epsilon^2/(4\pi \hbar^2 v_g^2)$ \cite{Jeong2011}. When considering a linear dispersion and an energy-independent $\lambda$, the McKelvey-Shockley flux equations need to be solved only once instead of at each energy (see Appendix \ref{simple_mcflux}). For numerical details see Appendix \ref{num_details}.

Fig. \ref{fig3} presents the normalized temperature profile ($T(x,t)-T_R/(T_L-T_R)$) versus normalized position $x/L$ for diamond films of length $L=0.1\,{\rm \mu m}$ (a), 1$\,{\rm \mu m}$ (c) and 10$\,{\rm \mu m}$ (e). Lines are solutions of the McKelvey-Shockley flux method and symbols are results of the full phonon BTE taken from \cite{Joshi1993}. The temperature profiles are plotted at different normalized times $\tau = t / t_{\rm ball}$, where $t_{\rm ball} = L/v_g$ is the ballistic transit time for a phonon traveling at the group velocity $v_g$.

An important feature of the temperature profiles is the temperature discontinuities at the ends of the thermal conductor. These temperature jumps at the contacts ($\delta T_c$) are signatures of ballistic phonon transport \cite{Joshi1993,Chen2002,Maassen2014}, and do not arise in conventional solutions of the heat equation or the hyperbolic heat equation \cite{Joshi1993,Chen2002} (the HE and HHE would give normalized temperatures of 1 and 0 at the left and right contacts, respectively). $\delta T_c$ decreases as $L$ increases and transport shifts from ballistic ($\lambda>>L$) to diffusive ($\lambda<<L$). We previously showed that in steady-state $\delta T_c = \mathcal{T}\Delta T/2$ \cite{Maassen2014}, where $\mathcal{T}=\lambda/(\lambda+L)$ is the transmission coefficient describing the probability of a phonon traveling from one contact to the other (ballistic: $\mathcal{T}\rightarrow 1$, diffusive: $\mathcal{T}\rightarrow \lambda/L$). The agreement of $\delta T_c$ between our approach and the full phonon BTE, over varying $L$ and $\tau$, is good; especially considering the enormous reduction in complexity of the McKelvey-Shockley flux method. We reemphasize that our contacts are reflectionless, thus $\delta T_c$ is not the result of added phonon scattering at the interface -- it is the signature of ballistic transport.

\begin{figure}	
\includegraphics[width=8.5cm]{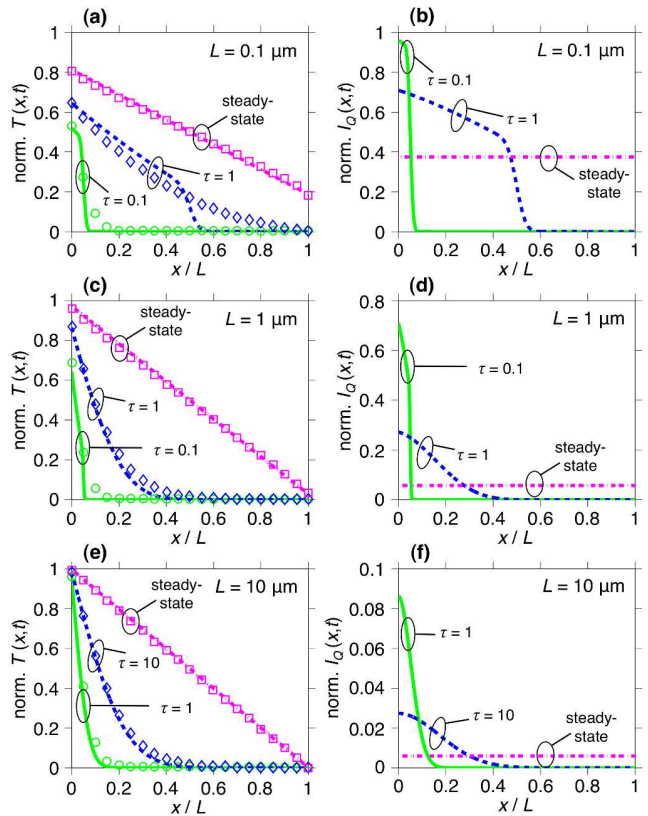}
\caption{(Color online) Normalized temperature profile $(T(x,t)-T_R)/(T_L-T_R)$ and normalized heat current $I_Q(x,t)/I_Q^{\rm ball}$ in diamond films versus normalized position $x/L$, where $I_Q^{\rm ball}$ is the steady-state ballistic heat current. Three film lengths are considered: $L$ = $0.1\,{\rm \mu m}$ (a)-(b), $1\,{\rm \mu m}$ (c)-(d), $10\,{\rm \mu m}$ (e)-(f). The results are plotted at different normalized times $\tau = t/t_{\rm ball}$, where $t_{\rm ball}=L/v_g$ is the ballistic transit time and $v_g$ is the group velocity. Lines are solutions of the McKelvey-Shockley flux method, and symbols correspond to the phonon BTE \cite{Joshi1993}.} 
\label{fig3}
\end{figure}

Another important characteristic observed in the temperature profiles is a finite-velocity propagation of phonons. This is most easily seen in the $L=0.1\,{\rm \mu m}$ film at short times $\tau=0.1$-1, where a wave-front behavior is clearly observed as phonons travel into the thermal conductor. A finite $\lambda$ makes the wave-front decay as it moves further into the material; `$+$' phonons scatter and become `$-$' phonons. For an isotropic phonon dispersion we have $v_x^+ = v_g/2$, thus on average phonons have traveled only $L/2$ at time $\tau = 1$, as shown in Fig. \ref{fig3}(a). The average time required by the phonon packet to traverse the film transitions from $(L/v_x^+)$ for ballistic transport to $(L^2/D_{\rm ph})$ for diffusive transport, where $D_{\rm ph}=v_x^+\lambda  /2$ is the phonon diffusion coefficient \cite{Maassen2014}.

The differences between the McKelvey-Shockley approach and the full BTE, observed in Fig. \ref{fig3}, are easily understood. In the case of $L=0.1\,{\rm \mu m}$, when transport is quasi-ballistic, the BTE shows a smoother temperature profile than McKelvey-Shockley. The temperature profile obtained with the BTE is smoother since it considers phonons with $v_x$ varying from 0 to $v_g$, however only a single, angle-averaged $v_x^+$ is used with McKelvey-Shockley leading to a more abrupt temperature distribution. A finer discretization in angle could be used, but that would just make the approach identical to the full BTE. The simple discretization into forward and reverse streams eliminates two degrees of freedom (a summation over $k$-points in the 3D Brillouin zone is replaced by an integration over energy) while retaining good accuracy under transient conditions and producing steady-state results that are identical to the BTE \cite{Maassen2014}.

In Fig. \ref{fig3} the normalized heat current $I_Q(x,t)/I_Q^{\rm ball}$ versus $x/L$ is presented for $L=0.1\,{\rm \mu m}$ (b), 1$\,{\rm \mu m}$ (d) and 10$\,{\rm \mu m}$ (f), where $I_Q^{\rm ball} = K^{\rm ball} \Delta T$ is the steady-state ballistic heat current and $K^{\rm ball} = C_V v_x^+/2$ is the ballistic thermal conductance \cite{Jeong2011}. Interestingly, at early times the heat current is much larger than that at steady-state. This happens because at very short times the `$+$' phonons injected at $x=0$ have not had the time to backscatter, leading to enhanced $I_Q$. The transient $I_Q$ eventually settles to a constant steady-state value, which corresponds to the transmission coefficient $\mathcal{T}=I_Q/I_Q^{\rm ball}$ \cite{Maassen2014}. In the absence of internal heat generation, a position-independent steady-state $I_Q$ is guaranteed with the McKelvey-Shockley flux method; by subtracting Eq.  (\ref{mk_flux2}) from Eq. (\ref{mk_flux1}) and using Eqns. (\ref{def_net_heat_curr})-(\ref{def_heat_den}), we obtain the energy balance equation 
\begin{align}
\frac{{\rm d}Q}{{\rm d}t} = - \frac{{\rm d}I_Q}{{\rm d}x}. \label{energy_bal_eq} 
\end{align}

\begin{figure}	
\includegraphics[width=8.5cm]{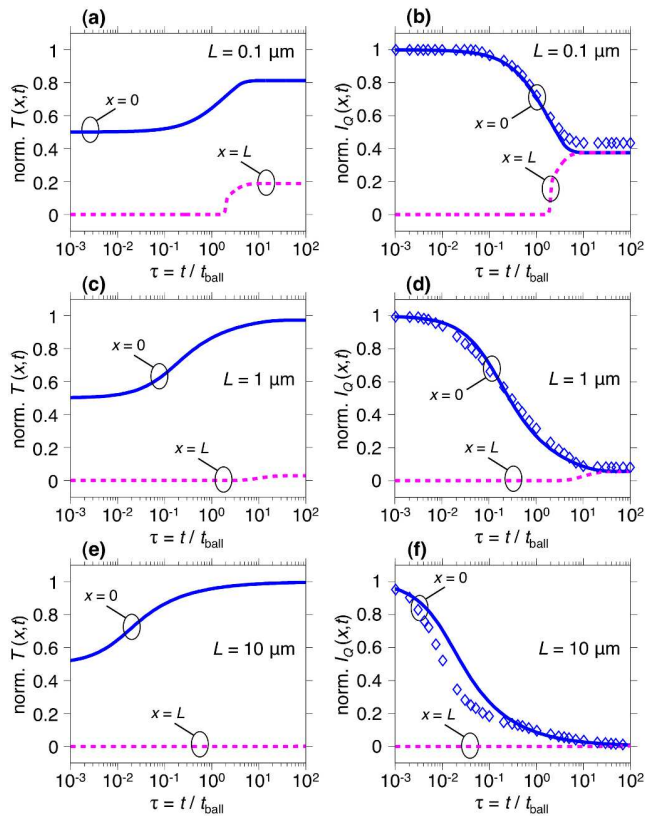}
\caption{(Color online) Normalized temperature profile $(T(x,t)-T_R)/(T_L-T_R)$ and normalized heat current $I_Q(x,t)/I_Q^{\rm ball}$ in diamond films versus normalized time $\tau = t/t_{\rm ball}$, where $I_Q^{\rm ball}$ is the steady-state ballistic heat current, $t_{\rm ball} = L/v_g$ is the ballistic transit time and $v_g$ is the group velocity. Three film lengths are considered: $L$ = $0.1\,{\rm \mu m}$ (a)-(b), $1\,{\rm \mu m}$ (c)-(d), $10\,{\rm \mu m}$ (e)-(f). Lines are solutions of the McKelvey-Shockley flux method, and symbols correspond to the phonon BTE (with $I_Q$ normalized to reproduce the McKelvey-Shockley flux value at $\tau=10^{-3}$) \cite{Joshi1993}.} 
\label{fig4}
\end{figure}

Fig. \ref{fig4} shows the normalized temperature profile and normalized heat current versus normalized time for $L=0.1\,{\rm \mu m}$ (a)-(b), 1$\,{\rm \mu m}$ (c)-(d) and 10$\,{\rm \mu m}$ (e)-(f). We plot the thermal response at the left and right ends of the thermal conductor, $x=0$ and $x=L$, respectively. At very short times the $x=0$ temperature is always close to 1/2, then increases and saturates. Similarly the $x=0$ heat current is close to 1 (the ballistic limit), then decreases and saturates. Why does this happen?

Shortly after $t=0^+$, the `+' states are filled with phonons injected from the left contact, while the `$-$' states are empty. From Eq. (\ref{def_temp}), normalized temperature is simply $(\delta T^+ + \delta T^-)/2$ (assuming $\Delta T=1\,{\rm K}$ for simplicity), where at early enough times at the left side $\delta T^+=1$ and $\delta T^-=0$, which averages to 1/2. This is, in fact, the temperature profile of a ballistic thermal conductor (temperature drop occurs only at the contacts) \cite{Chen2000}. Having only the `$+$' states filled means that the heat current is maximal and will at the ballistic limit, as shown in Fig. \ref{fig4}.

Importantly, a normalized temperature approaching 1/2 accompanied by a heat current near the ballistic limit are observed even when $L=10\,{\rm \mu m>>\lambda}$. This shows that, at early enough time, ballistic phonon effects are important even in ``diffusive" samples. Such phenomena is relevant for rapid time-resolved thermal characterization techniques, including time- and frequency-domain thermoreflectance measurements \cite{Koh2007,Regner2013,Wilson2014}.

In Fig. \ref{fig4}, we find the left side temperature increases simultaneously as the heat current decreases. How are these two observations related? It is possible to express the temperature jump at the contact, physically originating from nonequilibrium ballistic effects (filled `$+$' states and empty `$-$' states), in terms of a contact resistance (see Appendix \ref{contact_resistance}):
 \begin{align}
\delta T_c(0,t) = \frac{R_{\rm th}^{\rm ball}\,I_Q(0,t)}{2}, \label{temp_jump}
\end{align}
where $R_{\rm th}^{\rm ball}=1/K^{\rm ball}=2/C_Vv_x^+$ is the ballistic thermal resistance, and the factor of two comes from dividing the resistance over the left and right contacts. From Eq. (\ref{temp_jump}), we find that the temperature discontinuity is proportional to the heat current at that point, which is exactly what we observe when comparing temperature and heat current in Fig. \ref{fig4}.

The results presented in Figs. \ref{fig3} and \ref{fig4} show that, for this simple model problem, a full treatment of the phonon BTE agrees surprisingly well with a very simple phonon BTE of the form of Eqns. (\ref{mk_flux1})-(\ref{mk_flux2}).

\subsection{Silicon film: full phonon dispersion and energy-dependent mean-free-path}
\label{sec:silicon}
\begin{figure}	
\includegraphics[width=8.5cm]{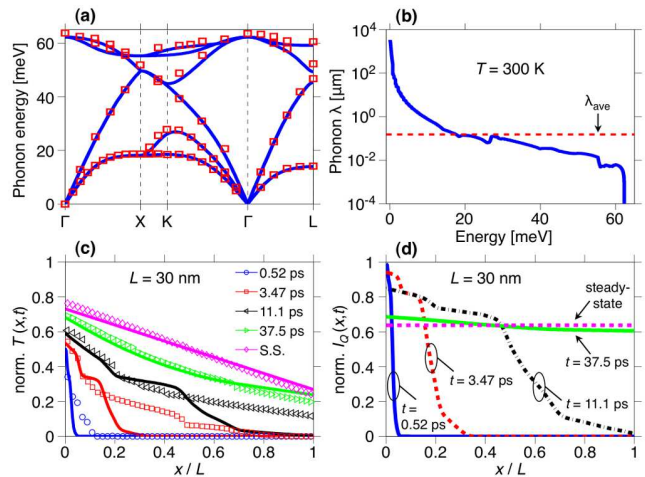}
\caption{(Color online) (a) Bulk Si phonon dispersion calculated from first principles. Symbols are measured data \cite{Nilsson1972}. (b) MFP distribution versus energy, including boundary, defect and Umklapp scatterings. Dashed line is the calculated average bulk $\lambda$ equal to $151\,{\rm nm}$. (c) Normalized temperature profile $(T(x,t)-T_R)/(T_L-T_R)$ and (d) normalized heat current $I_Q(x,t)/I_Q^{\rm ball}$ in a $L=30\,{\rm nm}$ Si film versus normalized position $x/L$, where $I_Q^{\rm ball}$ is the steady-state ballistic heat current. Lines are solutions of the McKelvey-Shockley flux method, and symbols correspond to the phonon BTE \cite{Escobar2008}.} 
\label{fig5}
\end{figure}

Up to this point, we have considered a diamond film using a linear phonon dispersion and energy-independent $\lambda$. For the vast majority of situations, however, accurate quantitative modeling of phonon transport requires full phonon dispersions and an energy-dependent $\lambda(\epsilon)$. Here, we calculate the transient thermal transport properties of a silicon film, using the phonon band structure of bulk Si extracted from first principles (Fig. \ref{fig5}(a)) and an energy-dependent $\lambda(\epsilon)$ including boundary, defect and phonon-phonon scatterings (Fig. \ref{fig5}(b)), calibrated to experimental data (details found in \cite{Maassen2014}). The total (energy-integrated) heat current is obtained from the energy-dependent forward/backward heat fluxes $I_Q^{\pm}(\epsilon)$ using Eq. (\ref{def_heat_curr_tot}). See Appendix \ref{temp_energy-dep} for how to calculate the total temperature profile from $I_Q^{\pm}(\epsilon)$.

Fig. \ref{fig5} presents (c) the normalized temperature and (d) the normalized heat current of a $L=30\,{\rm nm}$ Si film versus $x/L$ at different times $t$. The symbols correspond to results of the full phonon BTE taken from Ref. \cite{Escobar2008}, where a full phonon dispersion and energy-dependent MFP was also used. The overall agreement between the McKelvey-Shockley flux method and the full BTE is quite good; particularly given the greatly reduced computational effort of our approach.

At short times, McKelvey-Shockley flux shows a slower propagation speed since we use a single angle-averaged $x$-projected velocity $v_x^+(\epsilon)<v_g(\epsilon)$, where $v_g(\epsilon)$ is the group velocity. Our temperature discontinuities compare very well to those of the full BTE, except at the right side at short times (as just discussed). Compared to the case of diamond, we observe more features in the temperature and heat current profiles, because different phonons can travel at different velocities and scatter at different rates governed by the energy-dependent parameters $v_x^+(\epsilon)$ and $\lambda(\epsilon)$. We note in passing, that using the average bulk $\langle \lambda \rangle$ leads to significantly different results compared to using the energy-dependent $\lambda(\epsilon)$.

In summary, the McKelvey-Shockley flux method captures the general trends and essential physics (i.e. temperature jumps and finite-velocity propagation of heat) of the phonon BTE, and can be a simple alternative for modeling heat transport.

\section{Cattaneo Equation and the Hyperbolic Heat Equation} 
\label{sec:hhe}
We have shown that the McKelvey-Shockley flux method can treat heat transport including ballistic effects, finite-velocity propagation and can provide reasonable accuracy to the phonon BTE. Here, we demonstrate how the McKelvey-Shockley equations can be recast into familiar diffusion equations, {\it without approximation}.

By adding the two McKelvey-Shockley flux equations (Eqns. (\ref{mk_flux1})-(\ref{mk_flux2}), with $F^{\pm}$ replaced by $I_Q^{\pm}$), and using the definitions for the net heat current (Eq. (\ref{def_net_heat_curr})) and heat density (Eq. (\ref{def_heat_den})), we find:
\begin{align}
I_Q + \tau_Q\,\frac{\partial I_Q}{\partial t} &= - D_{\rm ph} \,\frac{\partial Q}{\partial x}, \label{cattaneo1} \\
&= - \kappa \,\frac{\partial T}{\partial x}, \label{cattaneo2}
\end{align}
where $\tau_Q = \lambda/(2v_x^+)$ is the heat relaxation time, $D_{\rm ph}=v_x^+\lambda / 2$ is the phonon diffusion coefficient and $\kappa = C_V D_{\rm ph}=C_V v_x^+ \lambda / 2$ is the bulk thermal conductivity (this expression for $\kappa$ is equal to the traditional relation $C_V v_g l/3$, where $l$ is the mean-free-path \cite{Maassen2014}). Eq. (\ref{cattaneo2}) is simply the Cattaneo equation \cite{Joshi1993,Chen2002}. Combining the Cattaneo equation with the energy balance equation (Eq. (\ref{energy_bal_eq})), we obtain:
\begin{align}
\tau_Q\,\frac{\partial^2 T}{\partial t^2} + \frac{\partial T}{\partial t}  &=  D_{\rm ph} \,\frac{\partial^2 T}{\partial x^2}, \label{hhe}
\end{align}
where we assume the material parameters are $x$-independent. Eq. (\ref{hhe}) is the well-known hyperbolic heat equation, derived from the McKelvey-Shockley flux equations {\it without making any assumption on $L$ relative to $\lambda$}. This indicates that ballistic phonon effects, which are present in the McKelvey-Shockley flux approach, are also captured by the HHE. Hence, all the results presented in this work are in fact simply solutions of the HHE. 

The key to capturing ballistic transport within the HHE is to use the correct physical boundary conditions. We previously demonstrated that with the proper boundary conditions Fourier's law and the HE can treat steady-state heat transport on {\it all} length scales, and accurately reproduce results of the phonon BTE \cite{Maassen2014}. The fact that the first order Boltzmann equation can be rewritten exactly as a second order equation is well known in neutron transport \cite{DuderstadtBook,LewisBook}.

Traditionally, one solves the HHE by using the contact temperatures, $T_L$ and $T_R$, as the boundary conditions. As we showed earlier, the injected heat fluxes, $I_Q^+(0^+,t)$ and $I_Q^-(L^-,t)$, are the correct physical boundary conditions. Specifying one flux component at only one end of the thermal conductor ($I_Q^+$ at $x=0^+$, and $I_Q^-$ at $x=L^-$), and not both ends, is key to capturing the nonequilibrium nature of ballistic transport.

Although the McKelvey-Shockley flux equations are {\it exactly} equivalent to the Cattaneo/hyperbolic heat equation, we find it most convenient to solve the former instead of the HHE. If one desired to use the HHE (in the case of ideal contacts), ballistic contact resistances could be introduced to effectively capture $\delta T_c$ (Eq. (\ref{temp_jump}) and Appendix \ref{contact_resistance}).

The HHE has previously been derived from the BTE \cite{Majumdar1993} by assuming local thermodynamic equilibrium at each $x$ (we showed earlier that this is the condition of diffusive transport), while we found no such approximation to be necessary. The ballistic-diffusive equations of Chen \cite{Chen2002} based on the BTE also yielded an expression similar to the HHE, with additional terms due to the ballistic phonons. We find that not only is temperature a solution of the HHE, but so are net heat current $I_Q$ and heat density $Q$ which can replace $T$ in Eq. (\ref{hhe}). The HHE derived here suggests that heat travels at a velocity of $\sqrt{D_{\rm ph}/\tau_Q}=v_x^+$, as expected from the McKelvey-Shockley flux equations. Traditionally, the HHE is solved assuming heat propagates (in an isotropic material) at a velocity $v_g/\sqrt{3}$. However in the case of McKelvey-Shockley (which can be derived from the phonon BTE), we find an average propagation velocity of $v_x^+ = v_g/2$. Lastly, we note that for an isotropic phonon dispersion, we have $\lambda = (4/3)\,l=(4/3)\,v_g\,\tau$ \cite{Jeong2010}, where $l$ is the mean-free-path and $\tau$ is the phonon scattering time. Using this relation in $\tau_Q = \lambda/(2v_x^+)$, we find $\tau_Q = (4/3)\,\tau$, indicating that the time required for a phonon to scatter from a forward-moving state to a backward-moving state is $4/3$ larger than the scattering time.

\section{Summary}
\label{sec:summary}
In conclusion, we have shown how to use the McKelvey-Shockley flux method to treat transient heat transport. By analyzing the transient response of diamond and silicon films, we demonstrated that this approach {\it (i)} captures ballistic phonon effects, such as temperature jumps at the boundaries, {\it (ii)} finite-velocity heat propagation, and {\it (iii)} can easily support full phonon dispersions and energy-dependent MFPs for detailed modeling. Our results show surprisingly good agreement with the phonon BTE, while requiring substantially less computational effort. Interestingly, we found that the McKelvey-Shockley flux equations can be rewritten, without approximation, as the Cattaneo and hyperbolic heat equations (HHE). The key to capturing ballistic effects in these diffusive equations is to use the injected heat fluxes as the boundary conditions.

Several physical insights are gained: {\it (i)} The temperature jumps at the contacts can be simply captured by including a interface resistance equal half the ballistic thermal resistance (a material property). The origin of the temperature jumps is related to the non-equilibrium nature of ballistic transport, and is not the result of enhanced scattering at the contacts. {\it (ii)} We demonstrate that at early enough times, phonon transport can approach the ballistic limit in samples of any length, which is relevant for time-resolved thermal experiments. {\it (iii)} We show that the HHE is applicable on all length and time scales, which provides an explanation for why diffusion equations work well in situations where ballistic effects are present.

The approach presented in this work is conditioned by two approximations. We assume a small applied temperature difference, and we use an angle-averaged $x$-projected phonon velocity at each energy. The latter results in wave-front transient heat propagation, rather than a smooth one expected in practice due to the wide range of $v_x$ from phonons traveling at all angles.

Given that this approach is strongly connected to fundamental physics (it is just a simple BTE derived directly from the full BTE), it represents a promising framework for analyzing experiments and modeling structures on all length and time scales. Finally, this work suggests that existing heat transfer simulation tools, based on traditional diffusive equations, might be modified to capture ballistic heat transport.

\acknowledgements
This work was supported in part by DARPA MESO (Grant N66001-11-1-4107) and through the NCN-NEEDS program, which is funded by the National Science Foundation, contract 1227020-EEC, and the Semiconductor Research Corporation. J.M. acknowledges financial support from NSERC of Canada.

\appendix

\section{Gray approximation of energy-independent $v_x^+$ and $\lambda$}
\label{simple_mcflux}
In the case of an energy-independent $v_x^+$ (i.e. linear phonon dispersion) and $\lambda$, the McKelvey-Shockley flux equations can be rewritten directly in terms of the energy-integrated total heat current ($I_Q^{\rm tot}$), heat density ($Q^{\rm tot}$) and temperature $\delta T^{\rm tot}$. By integrating the McKelvey-Shockley flux equations (Eqns. (\ref{mk_flux1})-(\ref{mk_flux2})) over energy, we find that $I_Q^{\pm}$ can be directly replaced with $I_Q^{\pm,{\rm tot}}=\int_0^{\infty}I_Q^{\pm}(\epsilon)\,{\rm d}\epsilon$: 
\begin{align}
\frac{1}{v_x^+}\frac{{\rm d} I_Q^{+,{\rm tot}}}{{\rm d} t} + \frac{{\rm d} I_Q^{+,{\rm tot}}}{{\rm d} x} &= -\frac{I_Q^{+,{\rm tot}}}{\lambda}+\frac{I_Q^{-,{\rm tot}}}{\lambda}, \label{mk_IQ1} \\
-\frac{1}{v_x^+}\frac{{\rm d} I_Q^{-,{\rm tot}}}{{\rm d} t} + \frac{{\rm d} I_Q^{-,{\rm tot}}}{{\rm d} x} &= -\frac{I_Q^{+,{\rm tot}}}{\lambda}+\frac{I_Q^{-,{\rm tot}}}{\lambda}. \label{mk_IQ2}
\end{align}
The boundary conditions become $I_Q^{+,{\rm tot}}(0^+,t)=\int_0^{\infty}I_Q^{+}(0^+,t,\epsilon)\,{\rm d}\epsilon$ and $I_Q^{-,{\rm tot}}(L^-,t)=\int_0^{\infty}I_Q^{-}(L^-,t,\epsilon)\,{\rm d}\epsilon$. By using the definition of total heat density (Eq. (\ref{def_heat_den}), and Eq. (\ref{def_heat_den_tot})), we have $Q^{\pm,{\rm tot}}=I_Q^{\pm,{\rm tot}}/v_x^+$. Hence, $I_Q^{\pm,{\rm tot}}$ in Eqns. (\ref{mk_IQ1})-(\ref{mk_IQ2}) can be replaced with $Q^{\pm,{\rm tot}}$: 
\begin{align}
\frac{1}{v_x^+}\frac{{\rm d} Q^{+,{\rm tot}}}{{\rm d} t} + \frac{{\rm d} Q^{+,{\rm tot}}}{{\rm d} x} &= -\frac{Q^{+,{\rm tot}}}{\lambda}+\frac{Q^{-,{\rm tot}}}{\lambda}, \label{mk_Q1} \\
-\frac{1}{v_x^+}\frac{{\rm d} Q^{-,{\rm tot}}}{{\rm d} t} + \frac{{\rm d} Q^{-,{\rm tot}}}{{\rm d} x} &= -\frac{Q^{+,{\rm tot}}}{\lambda}+\frac{Q^{-,{\rm tot}}}{\lambda}. \label{mk_Q2}
\end{align}
The boundary conditions for total heat density are $Q^{+,{\rm tot}}(0^+,t)=v_x^+\,I_Q^{+,{\rm tot}}(0^+,t)$ and $Q^{-,{\rm tot}}(L^-,t)=v_x^+\,I_Q^{-,{\rm tot}}(L^-,t)$. Finally, using the following properties ${\rm d}Q^{\rm tot}/{\rm d}x={\rm d}(\delta Q^{\rm tot})/{\rm d}x$, ${\rm d}Q^{\rm tot}/{\rm d}t={\rm d}(\delta Q^{\rm tot})/{\rm d}t$, $Q^+-Q^- = \delta Q^+ - \delta Q^-$ and $\delta Q^{\pm,{\rm tot}}=C_V^{\rm tot}\,\delta T^{\pm,{\rm tot}}$, Eqns. (\ref{mk_Q1})-(\ref{mk_Q2}) can be transformed into:
\begin{align}
\frac{1}{v_x^+}\frac{{\rm d} (\delta T^{+,{\rm tot}})}{{\rm d} t} + \frac{{\rm d} (\delta T^{+,{\rm tot}})}{{\rm d} x} &= -\frac{\delta T^{+,{\rm tot}}}{\lambda}+\frac{\delta T^{-,{\rm tot}}}{\lambda}, \label{mk_T1} \\
-\frac{1}{v_x^+}\frac{{\rm d} (\delta T^{-,{\rm tot}})}{{\rm d} t} + \frac{{\rm d} (\delta T^{-,{\rm tot}})}{{\rm d} x} &= -\frac{\delta T^{+,{\rm tot}}}{\lambda}+\frac{\delta T^{-,{\rm tot}}}{\lambda}, \label{mk_T2}
\end{align}
where $C_V^{\rm tot} = \int_0^{\infty}C_V(\epsilon)\,{\rm d}\epsilon$ and $\delta T^{\pm,{\rm tot}}=\int_0^{\infty} C_V(\epsilon)\delta T^{\pm}(\epsilon)\,{\rm d}\epsilon/C_V^{\rm tot}$. The boundary conditions for temperature are $\delta T^{+,{\rm tot}}(0^+,t)=T_L-T_{\rm ref}$ and $\delta T^{-,{\rm tot}}(L^-,t)=T_R-T_{\rm ref}$. 

In the case of an energy-independent $v_x^+$ and $\lambda$, the McKelvey-Shockley flux equations do not need to be evaluated at each energy $\epsilon$,  and can be simply solved once to extract $I_Q^{\pm,{\rm tot}}$, $Q^{\pm,{\rm tot}}$ and $\delta T^{\pm,{\rm tot}}$, thus significantly simplifying the computational effort.

\section{Numerical Solution of the McKelvey-Shockley Flux Method}
\label{num_details}
The coupled partial differential equations, that are the McKelvey-Shockley Flux equations (Eqns. (\ref{mk_flux1})-(\ref{mk_flux2})), are numerically solved using an explicit marching scheme for the spatial and temporal derivatives. Given the boundary conditions, it is natural to use a backward differencing in space for the forward-moving flux (Eq. (\ref{mk_flux1})) and a forward differencing in space for the backward-moving flux (Eq. (\ref{mk_flux2})). A forward differencing in time was chosen with both equations, using a time step $\Delta t  < [\Delta x \cdot \lambda  / (\Delta x + \lambda)] /v_x^+$ to ensure stability. A spatial grid resolution of $\Delta (x/L) \le 0.002$ was found to provide good accuracy.

\section{Temperature jumps and ballistic thermal resistance}
\label{contact_resistance}
The left side temperature jump is defined as $\delta T_c(0,t)=T_L - T(0^+,t)$. Noting that $T_L=T_R + \delta T^+(0^+,t)$, since in the contacts the forward/backward temperatures are equal, and using Eq. (\ref{def_temp}) we find
\begin{align}
\delta T_c(0,t) &= \frac{\left[\delta T^+(0^+,t) - \delta T^-(0^+,t)\right]}{2}. \label{temp_jump1}
\end{align}
The net heat current is $I_Q(x,t) = \delta I_Q^+(x,t) - \delta I_Q^-(x,t)$, which combined with Eq. (\ref{def_temp_pm}) gives
\begin{align}
I_Q(x,t) &= K^{\rm ball}\left[\delta T^+(x,t) - \delta T^-(x,t)\right], \label{heat_curr_temp}
\end{align}
where $K^{\rm ball} = C_V v_x^+/2$ is the ballistic thermal conductance. Combining Eq. (\ref{temp_jump1}) and Eq. (\ref{heat_curr_temp}), we obtain:
\begin{align}
\delta T_c(0,t) = \frac{R_{\rm th}^{\rm ball}\,I_Q(0,t)}{2}, \label{temp_jump2}
\end{align}
where $R_{\rm th}^{\rm ball}=1/K^{\rm ball}$ is the ballistic thermal resistance. Thus, the temperature jump can be effectively modeled as a contact resistance equal to the ballistic thermal resistance divided by two (for two contacts). An identical expression can be derived for the temperature jump at the right side ($\delta T_c(L,t)$).

\section{Energy-averaged temperature profile}
\label{temp_energy-dep}
Once the energy-dependent forward/backward heat currents $I_Q^{\pm}(\epsilon)$ are calculated by solving the McKelvey-Shockley flux equations (Eqns. (\ref{mk_flux1})-(\ref{mk_flux2})), the energy-dependent temperature profile $T(x,t,\epsilon)$ is obtained using Eqns. (\ref{def_temp})-(\ref{def_temp_pm}). The total (energy-integrated) temperature is determined using: 
\begin{align}
T^{\rm tot}(x,t) &= \frac{\int_0^{\infty} T(x,t,\epsilon)\,C_V(\epsilon)\,{\rm d}\epsilon}{\int_0^{\infty} C_V(\epsilon)\,{\rm d}\epsilon},
\end{align}
where $C_V(\epsilon)=\epsilon \,\{2 M(\epsilon)/h \,v_x^+(\epsilon)\} [\partial f_{\rm BE}(\epsilon)/\partial T]$ and $C_V^{\rm tot} = \int_0^{\infty}C_V(\epsilon)\,{\rm d}\epsilon$ is the total bulk heat capacity.


\begin{thebibliography}{00}
%
%
\bibitem{Cahill2014} D. G. Cahill, P. V. Braun, G. Chen, D. R. Clarke, S. Fan, K. E. Goodson, P. Keblinski, W. P. King, G. D. Mahan, A. Majumdar, H. J. Maris, S. R. Phillpot, E. Pop and L. Shi, Appl. Phys. Rev. {\bf 1}, 011305 (2014).
%
%
\bibitem{Koh2007} Y. K. Koh and D. G. Cahill, Phys. Rev. B {\bf 76}, 075207 (2007).
%
%
\bibitem{Regner2013} K. T. Regner, D. P. Sellan, Z. Su, C. H. Amon, A. J. H. McGaughey and J. A. Malen, Nature Comm. {\bf 4}, 1640 (2013).
%
%
\bibitem{Wilson2014} R. B. Wilson and D. G. Cahill, Nature Comm. {\bf 5}, 5075 (2014).
%
%
\bibitem{Minnich2011} A. J. Minnich, J. A. Johnson, A. J. Schmidt, K. Esfarjani, M. S. Dresselhaus, K. A. Nelson and G. Chen, Phys. Rev. Lett. {\bf 107}, 095901 (2011).
%
%
\bibitem{Sellan2010c} D. P. Sellan, J. E. Turney, A. J. H. McGaughey and C. H. Amon, J. Appl. Phys. {\bf 108}, 113524 (2010).
%
%
\bibitem{Bae2013} M.-H. Bae, Z. Li, Z. Aksamija, P. N. Martin, F. Xiong, Z.-Y. Ong, I. Knezevic and E. Pop, Nature Comm. {\bf 4}, 1734 (2013).
%
%
\bibitem{Esfarjani2011} K. Esfarjani, G. Chen and H. T. Stokes, Phys. Rev. B {\bf 84}, 085204 (2011).
%
%
\bibitem{Yang2013} F. Yang and C. Dames, Phys. Rev. B {\bf 87}, 035437 (2013).
%
%
\bibitem{Joshi1993} A. A. Joshi and A. Majumdar, J. Appl. Phys. {\bf 74}, 31 (1993).
%
%
\bibitem{Chen2002} G. Chen, J. Heat Transfer {\bf 320}, 124 (2002).
%
%
\bibitem{ChenBook} G. Chen, {\it Nanoscale Energy Transport and Conversion: A Parallel Treatment of Electrons, Molecules, Phonons, and Photons} (Oxford University Press, 2005).
%
%
\bibitem{Mahan1988} G. D. Mahan and F. Claro, Phys. Rev. B {\bf 38}, 1963 (1988).
%
%
\bibitem{Claro1989} F. Claro and G. D. Mahan, J. Appl. Phys. {\bf 66}, 4213 (1989).
%
%
\bibitem{Majumdar1993} A. Majumdar, J. Heat Transfer {\bf 115}, 7 (1993).
%
%
\bibitem{Narumanchi2006} S. V. J. Narumanchi, J. Y. Murthy and C. H. Amon, Heat Mass Transfer {\bf 42}, 478 (2006).
%
%
\bibitem{Escobar2008} R. A. Escobar and C. H. Amon, J. Heat Transfer {\bf 130}, 092402 (2008).
%
%
\bibitem{Sellan2010a} D. P. Sellan, J. E. Turney, A. J. H. McGaughey and C. H. Amon, J. Appl. Phys. {\bf 108}, 113524 (2010).
%
%
\bibitem{Hua2014} C. Hua and A. J. Minnich, Phys. Rev. B {\bf 89}, 094302 (2014).
%
%
\bibitem{Regner2014} K. T. Regner, A. J. H. McGaughey and J. A. Malen, Phys. Rev. B {\bf 90}, 064302 (2014).
%
%
\bibitem{Vermeersch2014a} B. Vermeersch, J. Carrete, N. Mingo and A. Shakouri, arXiv:1406.7341 (2014).
%
%
\bibitem{Vermeersch2014b} B. Vermeersch, A. M. S. Mohammed, G. Pernot, Y. R. Koh and A. Shakouri, arXiv:1406.7342 (2014).
%
%
\bibitem{Gomes2006} C. J. Gomes, M. Madrid, J. V. Goicochea and C. H. Amon, J. Heat Transfer {\bf 128}, 1114 (2006).
%
%
\bibitem{Hu2009} J. Hu, X. Ruan and Y. P. Chen, Nano Lett. {\bf 9}, 2730 (2009).
%
%
\bibitem{Sellan2010b} D. P. Sellan, E. S. Landry, J. E. Turney, A. J. H. McGaughey and C. H. Amon, Phys. Rev. B {\bf 81}, 214305 (2010).
%
%
\bibitem{Sullivan2013} S. E. Sullivan, K.-H. Lin, S. Avdoshenko and A. Strachan, Appl. Phys. Lett. {\bf 103}, 243107 (2013).
%
%
\bibitem{Mazumder2001} S. Mazumder and A. Majumdar, J. Heat Transfer {\bf 123}, 749 (2001).
%
%
\bibitem{Lacroix2005} D. Lacroix, K. Joulain and D. Lemonnier, Phys. Rev. B {\bf 72}, 064305 (2005).
%
%
\bibitem{Mittal2010} A. Mittal and S. Mazumder, J. Heat Transfer {\bf 132}, 052402 (2010).
%
%
\bibitem{Alvarez2007} F. X. Alvarez and D. Jou, App. Phys. Lett. {\bf 90}, 083109 (2007).
%
%
\bibitem{Maznev2011} A. A. Maznev, J. A. Johnson and K. A. Nelson, Phys. Rev. B {\bf 84}, 195206 (2011).
%
%
\bibitem{Wilson2013} R. B. Wilson, J. P. Feser, G. T. Hohensee and D. G. Cahill, Phys. Rev. B {\bf 88}, 144305 (2013).
%
%
\bibitem{Maassen2014} J. Maassen and M. Lundstrom, arXiv:1408.1631 (2014).
%
%
\bibitem{Mckelvey1961} J. P. McKelvey, R. L. Longini and T. P. Brody, Phys. Rev. {\bf 123}, 51 (1961).
%
%
\bibitem{Shockley1962} W. Shockley, Phys. Rev. {\bf 125}, 1570 (1962).
%
%
\bibitem{Alam1995} M. A. Alam, S.-I. Tanaka and M. S. Lundstrom, Solid-State Electron. {\bf 38}, 177 (1995).
%
%
\bibitem{Jeong2010} C. Jeong, R. Kim, M. Luisier, S. Datta and M. Lundstrom, J. Appl. Phys. {\bf 107}, 023707 (2010).
%
%
\bibitem{footnote_velo_mfp} As shown in Ref. \cite{Jeong2010}, the definitions of mean-free-path for backscattering and average $x$-projected velocity are: \begin{align} 
\lambda(\epsilon) &= 2 \,\frac{\sum_{k,v_x>0} v_x^2(k) \tau(k)\delta(\epsilon-\epsilon_k)}{\sum_{k,v_x>0} v_x(k) \delta(\epsilon-\epsilon_k)}, \nonumber \\ v_x^+(\epsilon) &=\frac{\sum_{k,v_x>0}v_x(k)\,\delta(\epsilon-\epsilon_k)}{\sum_{k,v_x>0} \delta(\epsilon-\epsilon_k)}. \nonumber
\end{align}
%
%
\bibitem{Alam1993} M. A. Alam, M. A. Stettler and M. S. Lundstrom, Solid State Elec. {\bf 36}, 263 (1993).
%
%
\bibitem{Rhew2002} J.-H. Rhew and M. S. Lundstrom, J. Appl. Phys. {\bf 92}, 5196 (2002).
%
%
\bibitem{Jeong2011} C. Jeong, S. Datta and M. Lundstrom, J. Appl. Phys. {\bf 109}, 073718 (2011).
%
%
\bibitem{Maassen2013a} J. Maassen, C. Jeong, A. Baraskar, M. Rodwell and M. Lundstrom, Appl. Phys. Lett. {\bf 102}, 111605 (2013).
%
%
\bibitem{Maassen2013b} J. Maassen and M. Lundstrom, Appl. Phys. Lett. {\bf 102}, 093103 (2013).
%
%
\bibitem{footnote_fit_mfp} The reported $\lambda=596\,{\rm nm}$ in Ref. \cite{Majumdar1993} was previously used to obtain excellent agreement with the results of the steady-state phonon BTE \cite{Maassen2014}. However, using this $\lambda$, a satisfactory agreement is not found for the same results in Ref. \cite{Joshi1993}. By fitting the data, in this work, we obtain good agreement with $\lambda=60\,{\rm nm}$.
%
%
\bibitem{Chen2000} G. Chen, J. Nanopart. Res. {\bf 2}, 199 (2000).
%
%
\bibitem{Nilsson1972} G. Nilsson and G. Nelin, Phys. Rev. B {\bf 6}, 3777 (1972).
%
%
\bibitem{DuderstadtBook} J. J. Duderstadt and W. R. Martin,{\it Transport Theory} (Wiley-Interscience Publications, 1979).
%
%
\bibitem{LewisBook} E. E. Lewis and W. F. Jr. Miller, {\it Computational Methods of Neutron Transport} (Wiley, 1984).
%
%


\end{thebibliography}
\end{document}